\documentclass{aa}
\usepackage{graphicx}
\usepackage{times}

\begin{document}

\thesaurus{08
           (08.01.1; 
            08.01.3; 
            08.16.4; 
            08.09.2 IRAS 04296+3429)}

\title{The peculiar post--AGB supergiant IRAS\,04296+3429:
      \protect\\ optical spectroscopy and its spectral
      energy distribution}

\author{V.G.\,Klochkova$^1$,  R.\,Szczerba$^2$, V.E.\,Panchuk$^1$,
                K.\,Volk$^3$}

\offprints{V.G. Klochkova (valenta@sao.ru)}

\institute{Special Astrophysical Observatory,
           N.Arkhyz, 357147 Russia
\and
           N.\,Copernicus Astronomical Center, PL-87-100 Toru\'{n}, Poland 
\and
           The University of Calgary, Calgary, Alberta T2N 1N4, Canada
}

\date{Received ....; accepted .....}

\titlerunning{Klochkova et al.~Peculiar supergiant IRAS04296+3429}
\authorrunning{V. Klochkova et al.}

\maketitle

\begin{abstract}

 The  optical  spectrum  of  the  infrared source IRAS\,04296+3429
 (optical  counterpart  --  G0\,Ia  star, V\,=\,14.2) was obtained
 with the echelle spectrometer PFES at the prime focus of the 6\,m
 telescope. We discover  {\it emission} bands (0,0) and (0,1) of
 the Swan system of the ${\rm  C_2}$ molecule in the optical spectrum
 of IRAS\,04296+3429. Comparison  with the spectrum of the Hale-Bopp
 comet  leads  us to propose that in both cases the same mechanism
 (resonance  fluorescence)  is  responsible for the emission in the
 ${\rm C_2}$ molecular bands.

 Several  strong  absorption features  whose positions coincide
 with  known diffuse interstellar bands (DIBs) are revealed in the
 spectrum of IRAS\,04296+3429.

 The  infrared  spectrum  of  IRAS\,04296+3429  shows  the  famous
 21\,${\mu}$m  feature (Kwok et al.~1989), but this object has not
 been   observed   by  KAO  (Omont  et  al.~1995).  However,  like
 IRAS\,05113+1347,  IRAS\,05341+0852 and IRAS\,22223+4327 (Kwok et
 al.~1995,  Szczerba et al.~1996), our detailed modelling of its
 spectral  energy  distribution  suggested  that  this source also
 should  show the 30\,${\mu}$m band. In fact, {\it ISO} discovered
 a  broad,  relatively  strong  feature  around  30\,${\mu}$m  for
 IRAS\,04296+3429 (Szczerba et al.~1999).

 The  surface  chemical composition of the source IRAS\,04296+3429
 is  metal-deficient  (the averaged value of the abundances of the
 iron group elements Ti, V, Cr and Fe relative to the solar values
 is  ${\rm  [M/H]_{\odot}\,=\,-0.9}$) and has been considerably
 altered  during  the  evolution:  carbon,  nitrogen and s-process
 elements are overabundant relative to the metallicity.

 The  totality  of  physical  and  chemical parameters derived for
 IRAS\,04296+3429  confirms  a  relation  between  presence of the
 feature  at  21\,${\mu}$m  in  the spectrum of a carbon rich star
 and an excess of the s-process elements.

\end{abstract}

\section{Introduction}

 This  paper  is a continuation  of  our  series  of  papers on the
 investigation   of   stars  which  are  believed  to  be  in  the
 post--Asymptotic  Giant  Branch  (post--AGB)  stage  of evolution
 (Klochkova~1995;  Za\v{c}s  et  al.~1995,  1996;  Klochkova \&
 Panchuk~1996a, 1998;  Klochkova  et al.~1997a; Klochkova \& Mishenina~1998).
  The  post--AGB  stars  (hereafter  also  referred  to  as
 proto--planetary  nebulae  -- PPNe) being in the transition phase
 from  AGB  to  planetary nebulae offer an opportunity to study in
 detail  a chemical composition which has undergone changes due to
 nucleosynthesis  and  mixing  processes in the course of the stars
 evolution.   Here   we  present  new  results  for  the  peculiar
 supergiant with a large infrared excess IRAS\,04296+3429.

 On  the 12/25/60 $\mu$m colour--colour diagram from the IRAS data
 infrared   source   IRAS\,04296$+$3429  (hereafter  IRAS\,04296),
 associated  with  a  faint  carbon--rich star (Omont et al.~1993;
 Loup  et  al.~1993)  classified  as  type  G0\,Ia  by  Hrivnak et
 al.~(1994),  is  located  in  the  region  occupied  by planetary
 nebulae,  non--variable OH/IR stars, and proto--planetary nebulae
 (Iyengar \& Parthasarathy~1997). The object IRAS\,04296 belongs
 to  the  small  group  of  sources  which show a spectral feature
 around  21\,$\mu$m  (Kwok et al.~1989). This feature is seen only
 for  some  post--AGB  objects and has not been detected either in
 the  preceding (AGB) nor in succeeding (PN) evolution stage. Note
 that  a  search  for  new  21\,$\mu$m  emitters  by means of {\it ISO SWS}
 observations  among  candidates selected by Henning et al.~(1996)
 failed  to  give any detections (Henning, private communication).
 Using  a  medium-resolution  (3\,\r{AA})  optical  spectrum, Hrivnak~(1995)
 found   that   IRAS\,04296   is   a  strongly  reddenned
 (E(B-V)\,=\,1.3)   G-star   with   features  indicative  of  high
 luminosity,  with  molecular absorption   features
 of ${\rm C_2}$ and rarely   observed  features  of
 ${\rm  C_3}$ {\it of circumstellar origin}  and  quite  strong
 absorption  lines  of  s-process  elements (Ba, Sr, Y) indicating
 that  outer  layers  of  the  atmosphere of IRAS\,04296 have been
 enriched  by  products of nucleosynthesis. Therefore this star is
 very  well  suited  for the study of detailed chemical abundances
 which have been changed by the third dredge--up.
 Indeed, Decin et al.~(1998), using high resolution spectra,
 obtained the chemical abundance pattern for this object and
 concluded that its metal--deficient, carbon--rich atmosphere has
 large overabundances of s--process elements.

 In   Sect.\,2   we   describe   our  observational  material  for
 IRAS\,04296  and  discuss  its molecular features, comparing them
 with   the   corresponding  spectrum  of  the  Hale--Bopp  comet.
 Sect.\,3  is  devoted  to presentation of the main parameters and
 detailed  analysis  of  the  chemical  composition of IRAS\,04296
 derived  from  our  optical  spectra.  The  next section presents
 modelling  of  spectral  energy distribution for this source with
 the  aim  to  get insight into its physical parameters (mainly to
 determine  the stellar effective temperature which is crucial for
 the  chemical  composition  estimation).  Finally, in Sect.\,5 we
 discuss  the results obtained and compare them to the results for
 related objects.

\section{Optical spectrum of IRAS\,04296}

\subsection{Observations and spectra reduction}

 We   have   obtained   spectra   of   IRAS\,04296  with  the  CCD
 (1140\,{$\times$}\,1170  pixels)  equipped  echelle  spectrometer
 PFES  mounted at the prime focus of the 6\,m telescope of SAO RAS
 (Panchuk   et   al.~1998).   The   echelle-grating   with   ${\rm
 75\,gr/mm}$     and    with    a    blaze    angle    of    ${\rm
 \Theta\,=\,64.3\,^{o}}$  was  used.  A  diffracton  grating  with
 ${\rm  300\,gr/mm}$  was  used as the cross-disperser. The camera
 has f\,=\,140\,mm. The projected angular size of the input slit is 
 0.54\,arc sec.

 We  observed  IRAS\,04296  on  October 07, 1996 (JD2450363.6) and
 February  26,  1997  (JD2450506.3).  The  echelle  frames with 25
 echelle-orders     cover     the     spectral     region    ${\rm
 \lambda\lambda}$\,4420-8300\,\r{AA}.     The     average    spectral
 resolution  was  0.4\,\r{AA}.  The  signal-to-noise ratio was in the
 range 50--110 for different spectral orders.

 All  usual  procedures  needed for echelle-images reduction (bias
 subtraction,   cosmic  ray  removal,  optimal  order  extraction,
 rebinning)  were  made  using  the  ECHELLE  context of the MIDAS
 system.  A  Gaussian  function  approximation  was  made  for the
 measurement  of equivalent widths. The comparison spectrum source
 was an argon-filled thorium hollow-cathode lamp.

\begin{figure}
\resizebox{\hsize}{!}{\includegraphics{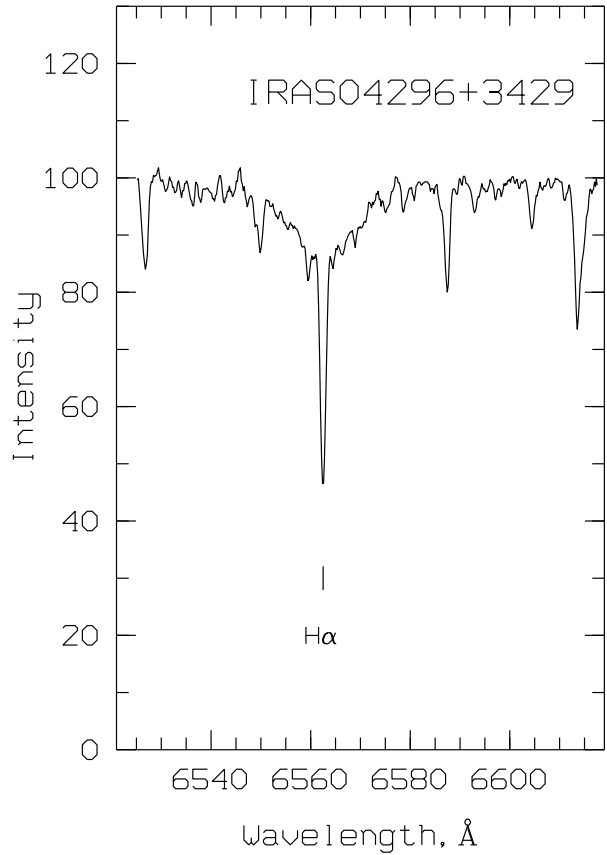}}
   \caption[]{
The IRAS\,04296+3429 spectrum near ${\rm H_{\alpha}}$.
}
\label{HI}
\end{figure}

 The  distinctive  features  of the optical spectrum of the source
 IRAS\,04296  are  a  peculiar  profile  of the ${\rm H_{\alpha}}$
 line (see Fig.\,\ref{HI}), molecular emission bands and very strong
 absorption lines of ionized atoms of s--process elements
 (Y, Zr, Ba, La, Ce, Pr, Nd). For example, the equivalent widths
 of Ba\,II lines (6141 and 6496\,\r{AA}) exceed 0.6\,\r{AA}.
%

\subsection{Emission molecular bands}

 Absorptional  bands  of  several molecules (${\rm C_2}$, CN, TiO,
 etc.)  are  often  present in the spectra of post-AGB stars (see,
 for   example,   Hrivnak~1995;  Bakker  et  al.~1997).  However,
 molecular  {\it  emission} features are only very rarely observed
 in  the  optical spectra of PPNe. One such example is RAFGL\,2688
 (the  Egg  Nebula)  for  which  Crampton  et  al.~(1975) observed
 emission  features  of  the  ${\rm  C_2}$  molecule  in  a medium
 resolution  spectrum.  On  the  other hand, it is well known that
 cometary nuclei spectra show prominent Swan band emission.

\begin{figure}
\resizebox{\hsize}{!}{\includegraphics{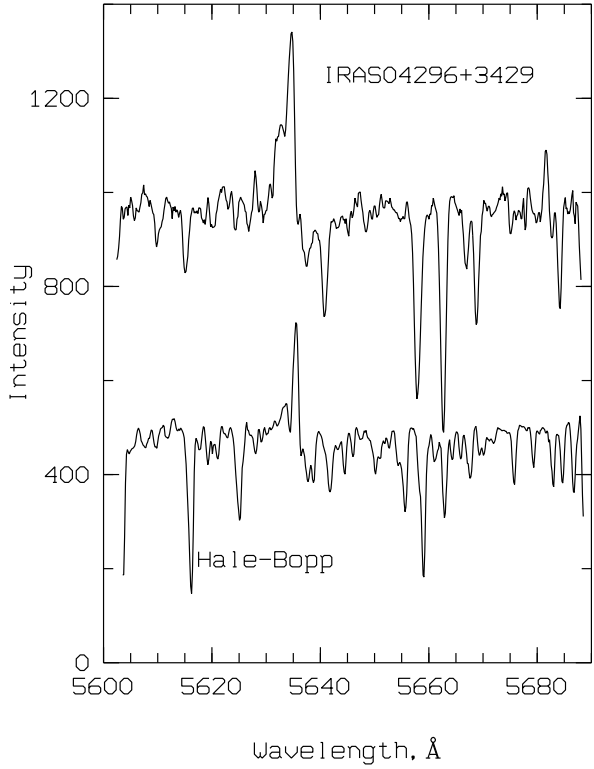}}
   \caption[]{
 Comparison  of  the  IRAS\,04296\,+\,3429
 ${\rm C_2}$ Swan band head (0;1) at ${\rm \lambda}$\,=\,5635\,\r{AA}
 with that for the Hale-Bopp comet nuclei
}
\label{Swan1}
\end{figure}

\begin{figure}
\resizebox{\hsize}{!}{\includegraphics{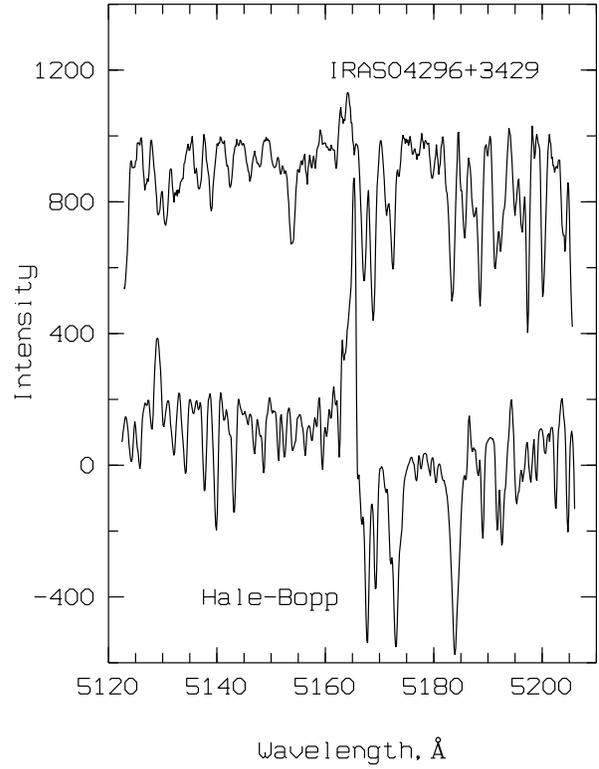}}
   \caption[]{
 The same as Fig.\,2 but for ${\rm C_2}$
 Swan band head (0;0) at ${\rm \lambda}$\,=\,5165\,\r{AA}
}
\label{Swan2}
\end{figure}

\begin{figure}
\resizebox{\hsize}{!}{\includegraphics{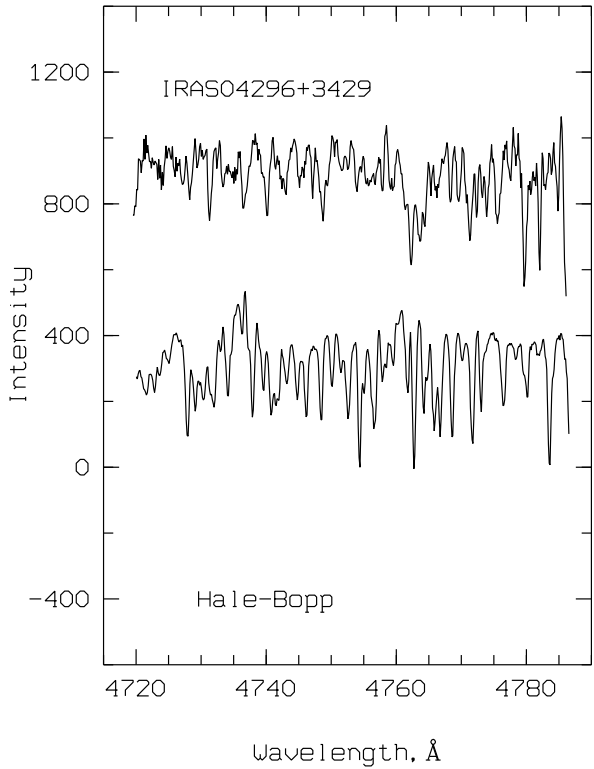}}
   \caption[]{
 The same as Fig.\,2 but for ${\rm C_2}$
 Swan band head (1;0) at ${\rm \lambda}$\,=\,4735\,\r{AA}}
\label{Swan3}
\end{figure}

 In  both  spectra  of  IRAS\,04296 we have discovered strong {\it
 emission}  in  the  (0;0)  and (0;1) bands of the Swan system of
 the  ${\rm  C_2}$  molecule. On Figs.~\ref{Swan1}--\ref{Swan3} we
 present   a   comparison  between  the  spectrum  of  IRAS\,04296
 (observed  on February, 26, 1997) and that of the Hale-Bopp comet
 (observed  on  March  30, 1997 with the same spectrometer) around
 bands    (0;1),    (0;0)    and    (1;0),    respectively.   From
 Figs.~\ref{Swan1}--\ref{Swan3}  it  is  clear  that emission band
 (1;0) at 4735\,\r{AA}\ is absent in  the spectrum of IRAS\,04296 while
 the bands  (0;1)  at  5635\,\r{AA}\ and (0;0) at 5165\,\r{AA}\ are reliably
 measured.
 Hrivnak~(1995) obtained the spectrum of IRAS\,04296
 inside the blue spectral region, 3872--4870\,\r{AA}\r{AA},
 therefore he could not observe emission features of ${\rm C_2}$ at
 5165 and 5635\,\r{AA}\r{AA}.
%
%
%

 To  understand  the  observed  ratios between different bands, we
 have  estimated  the  temperature function  for  monochromatic
 coefficient  of  absorption per molecule ($\sigma_{\lambda}$) for
 the  Swan  bands  in the ``just overlapping'' approximation (JOA,
 Golden~1967).    For   the   microturbulent   velocity   ${\rm
 \xi_t\,=\,7\,km/sec}$  this  approximation  works  well  near the
 band  heads. Values of $\sigma_{\lambda}$ for band heads of (0;0)
 at  5165\,\r{AA}\  and  (1;0) at 4735\,\r{AA}\ coincide within 0.2\,dex
 for     the     temperatures     range    3000--7000\,K,    while
 $\sigma_{\lambda}$   for   the   band   (0;1)  at  5635\,\r{AA}\  is
 systemically  lower  by about 0.6\,dex. Taking into account these
 relations  between  $\sigma_{\lambda}$'s for different band heads
 and since  we  do  not observe the 4735\,\r{AA} band in RAFGL\,2688
 and  IRAS\,04296,  we  can  conclude  that  it  is  impossible to
 describe  the  intensity ratios of ${\rm C_2}$ emission bands for
 these   objects   by   means   of  an equilibrium  vibrational
 temperature in the 3000--7000\,K range.

 To  explain  emission  bands intensities for comets the mechanism
 of resonance fluorescence has been proposed (Zanstra~1928, Swings~1941).  
In  that  case population of vibration--rotational levels
 for  the  molecule  is  described by the Boltzmann approximation,
 however the value of {\it T} in the exponent no longer has the
 meaning  of  equilibrium  temperature  but  it  is a distribution
 parameter  only.  We  suggest that the same mechanism could be
 responsible  for  the  observed  emission  bands  of IRAS\,04296.
 However,  it  is  clear from  Figs.~\ref{Swan1}--\ref{Swan3}
 that  there  are significant differences in the equivalent widths
 of  the  emission  bands  in the spectra of our supergiant and of
 the  Hale--Bopp  comet  nucleus.  They  could  be  explained by a
 difference  of  radiation  fluxes  which  illuminate  ${\rm C_2}$
 molecules  in  these  objects.  The  temperature  of  IRAS\,04296
 ($\rm T_{eff}$  around\,  6300\,K) is sufficiently higher than that for
 the  Sun,  therefore the band (1,0) at 4735\,\r{AA} for IRAS\,04296
 should  be  stronger  than  that  for the Hale-Bopp comet nuclei.
 However,  Fig.\,\ref{Swan3}  shows  the  opposite  behaviour.  It
 could  mean that radiation field of IRAS\,04296 which excites the
 ${\rm  C_2}$  molecules  is  strongly  reddened by matter located
 between  its  photosphere and the region which produces the ${\rm
 C_2}$ emission.

 Together  with  the  emission bands of the Swan system (Klochkova
 et  al.~1997b)  absorption  bands of the Phillips system (1:0),
 (2;0),  (3;0) have  been  revealed  in the spectrum of IRAS\,04296
 (Bakker  et  al.~1997).  Let  us try to explain this phenomenon
 within  the  resonance  fluorescence  mechanism  ordinary used to
 interpret comets' spectra. As a first approximation, we assume that
 the vibrational distribution corresponds to the
 effective  temperature  of the star  illuminating a circumstellar
 envelope  if  vibrational transitions in the low triplet state of
 a  homonuclear  molecule  are  strictly  forbidden. But even when
 interpreting  comets'  spectra such an approach appears to be too
 poor.  The intensity distributions for different systems of bands
 and for bands of individual systems of the resonance fluorescence
 of  the  ${\rm  C_2}$  molecule have been considered in papers by
 Krishna  Swamy \&  O'Dell~(1977,  1979,  1981). The intensities of
 bands  have been calculated taking into account the excitation of
 the   Swan,  Ballick-Ramsay  and  Fox-Herzberg  triplet  systems,
 Phillips  and Milliken singlet systems as well as singlet-triplet
 transitions  in  low  states.  It  has been shown, in particular,
 that  at  the  value of the moment of singlet-triplet transitions
 ${\rm  |R_e|^{2}=10^{-5}}$ and at the heliocentric distance of a
 comet  d=1\,a.u.  the  ratios  of  intensities of sequences ${\rm
 \Delta\nu=0,\,1,\,-1}$  in  the  Phillips system to the intensity
 of  sequence  ${\rm  \Delta\nu=0}$ of the Swan system is equal to
 0.094,  0.11  and  0.04,  correspondingly (Krishna Swamy \& O'Dell~1981). 
This agrees well with results of measurement of comets'
 spectra. Using these results of Krishna Swamy \& O'Dell~(1981), 
we may suppose that the intensity of main bands of the Swan
 system is ten times higher than that in the Phillips system.

 Now consider the case of IRAS\,04296. Let us add such an emission
 spectrum of the ${\rm  C_2}$  on the stellar continuum.
 In order to observe the emission bands of both the Swan and the
 Phillips systems over the continuum in such a combined spectrum,
 the stellar flux  at  ${\rm \lambda}$ = 5165\,\r{AA} 
 must be at least 10 times
 higher than  near ${\rm \lambda}$\,=\,7720\,\r{AA}.
 From Kurucz's~(1979) tables it follows that the ratio of the fluxes
 near these wavelenghts for the Sun (the emitter in the case of comets) is
 equal to ${\rm F_{\lambda 5175} / F_{\lambda 7750} = 1.5}$.
 For the model with $\rm T_{eff}$ = 6300\,K this ratio is equal to
 ${\rm F_{\lambda 5175} / F_{\lambda 7750} = 1.9}$.
 From the real spectral energy distrubution observed for
 IRAS\,04296 (Kwok~1993)  the ratio of the fluxes  is essentially smaller:
 ${\rm F_{\lambda 5165} / F_{\lambda 7720} = 0.1}$. Therefore,
 the  conditions  to  observe  the absorption  bands  of  the Phillips
 system  and  the emission  bands of the Swan system may arise
 inside the circumstellar envelope of IRAS\,04296.

\subsection{Radial velocity of the IRAS\,04296}

 We  measured  the  radial  velocity  ${\rm  V_r}$  using our best
 spectrum,   that   obtained  in  February,  1997.
 The average value  of the  radial  velocity  from numerous  metal
 lines   (${\rm V_r\,=\,-56.0\pm 0.8\,km/s}$) and from the
 $\rm  H_{\alpha}$ absorption line (${\rm   V_r\,=\,-54.5\,km/s}$) agree
 within  the accuracy of  measurement.
 This  radial  velocity  is consistent with a membership
 of an old population as suggested by the low
 metallicity  (see  Table\,1).  It  should be noted also that the
 value  of  radial  velocity  we  derived agrees with the value of
 ${\rm  V_r\,=\,-59\,km/s}$ which was given by Omont et al.~(1993)
 from  CO data and ${\rm V_r\,=\,-62\,km/s\pm 1.0\,km/s}$ by Decin
 et  al.~(1998)  from  optical spectrum. There is still no sign of
 an essential temporal variability of ${\rm V_r}$ for the object.

\subsection{Absorption bands identified with DIBs}

\begin{figure}
\resizebox{\hsize}{!}{\includegraphics{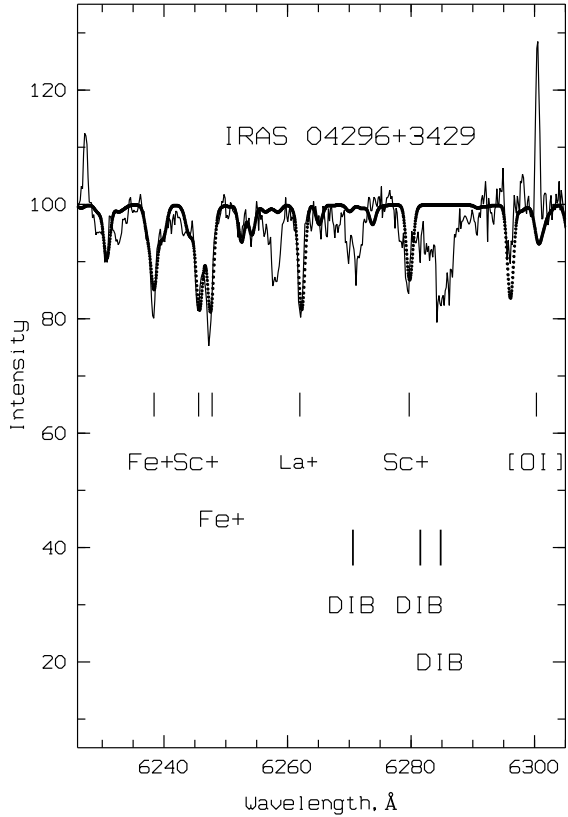}}
   \caption[]
 {The observed (line) and  synthetic (dots)
 spectra of the object IRAS\,04296\,+\,3429 near absorbtional features identified
 with the DIBs. The telliric spectrum was removed from the observed spectrum of
 the object}
\label{DIB}
\end{figure}

 From  comparison of observed and synthetic spectra of IRAS\,04296
 we  discovered  some  strong  absorptional features whose positions
 coincide   with   known   diffuse   interstellar   bands   (DIBs)
 (Jenniskens  et  al.~1994).
 In Fig.\,\ref{DIB} we illustrate the presence of DIB's by
 showing a spectral region of IRAS\,04296.  We have calculated the
 synthetic spectrum using the  code  STARSP  (Tsymbal~1995) and the
 atmospheric parameters and abundances  of chemical elements we here
 obtained. It should be  noted  that  for  such a comparison in the
 spectral range near
 ${\rm  \lambda}$  6270-6310\,\r{AA}  the  telluric spectrum has been
 removed from the observed spectrum.

 In   the  following  paper  we  plan  to  study  in  detail  such
 identified  with  DIB's  absorptions,  we  have  revealed  in the
 spectra     of     several    related    objects    (IRAS\,04296,
 IRAS\,23304+6147,  IRAS222223+4327),  here  we limited ourself by
 such short information.
%

\section{Determination of atmospheric parameters and calculation of
         chemical composition}

 For  understanding  of  an  object  at  an  advanced evolutionary
 stage,  it is very important to know its metallicity and detailed
 chemical  abundance  pattern.  Our echelle spectra provide such a
 possibility due to their large wavelengths coverage.

 To   study   the   chemical   composition,   we   have  used  the
 plane-parallel  homogeneous models generated by the MARCS program
 (Gustafsson  et  al.\,1975).  It  should  be noted, however, that
 unstable  and  very  extended atmospheres of supergiants probably
 require  more  advanced model atmospheres. Therefore, our results
 should  be  treated  as  only  preliminary  ones.  For a chemical
 composition  calculation  by  the  model  atmosphere  method, one
 needs  to  know  the values of the effective temperature ($\rm T_{eff}$),
 surface  gravity  (log\,g)  and  microturbulent  velocity  (${\rm
 \xi_t}$).  Determination  of $\rm T_{eff}$  is  problematic  even  for
 normal   supergiants   due  to  their  extended  atmospheres  and
 significant  non-LTE  effects.  In  the  case  of  so  peculiar a
 supergiant  as  IRAS\,04296, for which the energy distribution is
 strongly  distorted by interstellar and circumstellar extinction,
 determination  of $\rm T_{eff}$  is  the  most  difficult  problem. We
 cannot  use  for  this  purpose equivalent widths and profiles of
 H\,I  lines  (well  known  criteria of atmospheric conditions for
 normal  supergiants), since these lines are strongly distorted in
 the spectrum of IRAS\,04296 as seen in Fig.\,\ref{HI}.

 Therefore,   we   have   applied  the  spectroscopic  method  for
 temperature  determination  of IRAS\,04296, forcing the abundance
 derived  for  each  line  to  be  independent  of  the
 lower excitation potential (EP). We have estimated
 that $\rm T_{eff}$ = 6300\,K   with  an internal   uncertainty
 $\Delta$\,$\rm T_{eff}$ = 250\,K.  To  check the realiability of our
 determination  we  have modelled the spectral energy distribution
 for   this   source   (see  Sect.\,4)  and  got  a  very  similar
 temperature  near 6500\,K. The surface gravity log\,g\,=\,0.0 was
 estimated  through  the  ionization  balance  of the Fe\,I and Fe\,II
 abundances. The errors  on the parameter log\,g is determined by forcing a
 maximum  difference  between  ${\rm \epsilon\,(Fe\,I)}$ and ${\rm
 \epsilon\,(Fe\,II)}$  to  be  0.1\,dex (where here and hereafter,
 ${\rm  log\,\epsilon\,(X)\,=\,log\,N(X)-log\,N(H)}$).  It  should
 be  noted  that  the hydrogen abundance ${\rm log\,N(H)}$\,=\,12.
 Such  a  difference is achieved by varying the log\,g  value
 by ${\rm \pm 0.2}$  keeping  other  parameters
 ($\rm T_{eff}$  and  ${\rm  \xi_t}$) constant.
 The  microturbulent velocity value based on equivalent
 widths  (W)  of  Fe\,I  and  Fe\,II lines is quite high, equal to
 7\,km/s.  This  value  is determined with an uncertainty of ${\rm
 \pm  1.0\,km/s}$,  which  is  typical  for  F, G--supergiants.

\begin{figure}
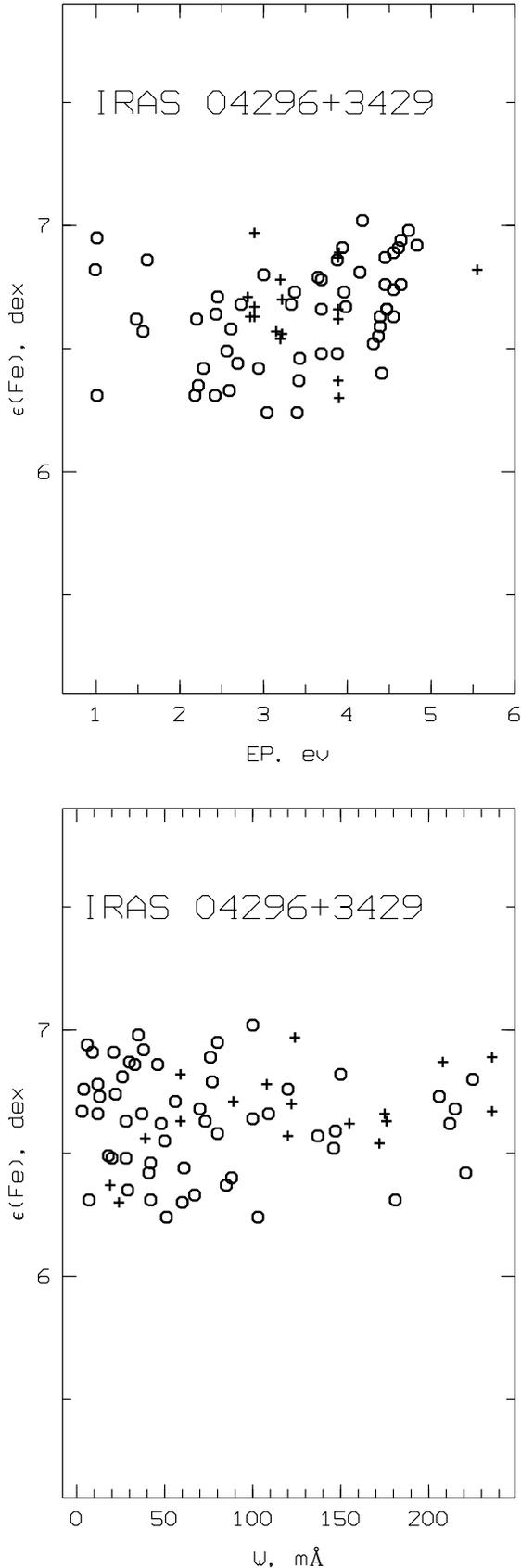

\resizebox{\hsize}{!}{\includegraphics{MS7673.f6a}}
\resizebox{\hsize}{!}{\includegraphics{MS7673.f6b}}
   \caption[]{
 {\it Upper:} iron abundance FeI (circles) and FeII (crosses) calculated
   for IRAS\,04296
 with model parameters ${\rm T_{eff}=6300\,K}$, ${\rm log\,g=0.0}$ and
 ${\rm \xi_t=7.0\,km/sec}$ using lines with different EP of a low level;
 {\it bottom:} the same as a function of the equivalent width W}
\label{Fe}
\end{figure}

 To  illustrate  the  choice  of  model  parameters for the object
 IRAS\,04296  in  the  Fig.\,\ref{Fe}  are  shown  the  excitation
 potential  --  abundance  diagram  and  the  equivalent  width --
 abundance  diagram  for  lines  of  neutral  (dots)  and  ionized
 (crosses)  iron atoms. As follows from this figure, there are no
 essential  dependences  for  values  considered.  The  large
 dispersion  is  mainly  explained  by  errors  of  measurement of
 equivalent  widths  of  weak  absorption  lines  for such a faint
 object   as  the  IRAS\,04296  (see,  for  example,  the  similar
 dispersion  on  the  Fig.\,1  in the paper by Decin et al.~(1998)
 for the brighter object IRAS\,22223+4327, V=9.7).

 We have checked the determination  of IRAS\,04296 model parameters
 using weaker FeI and FeII lines and concluded that the
 parameters are steady within the erorr box up to
 ${\rm W}$ = 100-150\,m\r{AA}. This can also be seen  from Fig.\,\ref{Fe}.
%

 It  is well known that the plane-parallel static model atmosphere
 method  does  not  give  correct  abundances  for high luminosity
 stars  (luminosity classes Ia, Ia+). The profiles of the spectral
 lines  observed  are  broadened by non-thermal mecha\-nisms whose
 influence  may be variable at different levels in the atmosphere.
 Therefore,  to obtain more reliable estimates of chemical element
 abundances  we  use  weak  lines with ${\rm W < }$250\,m\r{AA}.
 The average values of the equivalent widths ${\rm \overline{W}}$  we
 used for the abundances calculations  are also  given in Table\,1.
 Only the BaII abundance was  calculated  using 3 very strong lines:
 ${\rm   W(\lambda}$ 5853.67)\,=\,464\,m\r{AA},
 ${\rm W(\lambda}$ 6141.71)\,=\,679\,\,m\r{AA}
 and ${\rm W(\lambda}$ 6496.90)\,=\,738\,\,m\r{AA},
 because the weaker lines of this element were not available.
 In general, the weak lines formed in deeper atmospheric layers are
 more correctly described by the standard static model. The limitation
 of equivalent width of lines used to  ${\rm  W < }$250\,m\r{AA}
 significantly reduces the influence of uncertainty in the choice
 of ${\rm \xi_t}$. Note, however, that the main factor in the abundance
 errors for most species remains the uncertainty of the $\rm T_{eff}$ value.
 Therefore, we have checked our estimation  of $\rm T_{eff}$ by modelling of
 spectral energy distribution for IRAS\,04296.

 Computed  abundances  of  26  chemical  elements are presented in
 Table\,1.  In  the head of the Table\,1 parameters of the adopted
 model   atmosphere   are   shown.   The  dependence  of  chemical
 composition   determination   on   uncertanties   of   the  model
 atmosphere  parameters is discussed in Za\v{c}s et al.~(1995). In
 the second column of Table\,1 derived abundances are given as
 ${\rm log\,\epsilon\,(X)}$, while  in  the  third  column  estimated
 uncertainties of ${\rm \sigma\,=\,\Delta\,log\,\epsilon\,(X)}$ are shown.
 In the next column, the number of  spectral  lines  used for chemical
 composition calculation is indicated.

\begin{table*}
\caption[]{Model atmosphere parameters adopted and abundances of chemical
         elements. Here, n is number of
         lines used for calculation, ${\sigma}$ -
         the standard deviation, ${\rm \overline{W}}$ - the average equivalent
         width, in ${\rm m\r{AA}}$,  of lines  used for the content calculation}
\begin{tabular}{l|crrr|crrr}                                 
\hline
&&&&&&&& \\
&\multicolumn{4}{c|}{IRAS\,04296+3429} &\multicolumn{4}{c}{${\rm\alpha\,Per}$}\\
&\multicolumn{4}{c|}{$\rm T_{eff}$\,=\,6300\,K, log\,g\,=\,0.0, $\xi_t$\,=\,7.0\,km/s}
&\multicolumn{4}{c}{$\rm T_{eff}$\,=\,6500\,K, log\,g\,=\,1.2,$\xi_t$\,=\,4.7\,km/s}\\[5pt]
\hline
&\multicolumn{4}{c|}{}& \multicolumn{4}{}{}\\
 Element&${\rm log\,\epsilon (X)}$ & \hspace{0.6cm} ${\sigma}$
 & n  &${\rm \overline{W}}$& ${\rm log\,\epsilon (X)}$
 & \hspace{0.6cm}
 ${\sigma}$&n &${\rm \overline{W}}$  \\[3pt]
 \hline
 Li\,I        &  ${\rm \ge2.70}$ & &  1 & 32 &    &    & & \\
 C\,I         &    8.55 & 0.46  & 21 & 69 &8.16&0.14& 13 & 47 \\
 N\,I         &    7.96 & 0.10  &  4 & 99 &8.35&0.10&  4 & 127\\
 O\,I         &    8.22 & 0.05  &  3 & 26 &8.35&0.06&  4 & 23 \\
 Na\,I        &    5.91 & 0.24  &  3 & 68 &6.48&0.06&  4 & 48\\
 Mg\,I        &         &       &    &    &7.83&0.03&  2 & 56\\
 Mg\,II       &    8.08 & 0.03  &  2 &254 &    &    &    & \\
 Al\,I        &    6.66 & 0.14  &  3 & 68 &6.57&0.16&  4 & 32\\
 Si\.I        &    7.29 & 0.20  & 11 & 37 &7.68&0.16& 16 & 45\\
 Si\,II       &    6.97 &       &  1 & 22 &7.81&    &  1 & 278\\
 S\,I         &    6.80 & 0.21  &  7 & 30 &7.53&0.23&  2 & 187\\
 Ca\,I        &    5.71 & 0.30  & 19 & 98 &6.41&0.22& 14 & 122\\
 Sc\,II       &    2.51 & 0.28  & 10 &164 &2.72&0.07&  6 & 119\\
 Ti\,II       &    3.91 & 0.33  &  5 &184 &4.78&0.08&  4 & 47\\
 V\,II        &    3.26 & 0.28  &  4 & 26 &3.54&0.10&  4 & 22\\
 Cr\,II       &    4.94 & 0.28  & 10 &108 &5.54&0.12&  9 & 136\\
 Mn\,I        &         &       &    &    &5.25&0.09&  3 & 63\\
 Fe\,I        &    6.66 & 0.30  & 55 & 75 &7.48&0.21&111 & 59\\
 Fe\,II       &    6.65 & 0.22  & 19 &131 &7.51&0.09& 10 & 154\\
 Cu\,I        &    3.61 &       &  1 & 38 &4.66&    &  1 & 36\\
 Zn\,I        &    3.84 &       &  1 &  9 &    &    &    & \\
 Y\,II        &    2.60 & 0.14  &  2 &168 &2.20&0.40&  2 & 32\\
 Zr\,I        &         &       &    &    &3.38&0.10&  4 & 6\\
 Zr\,II       &    2.38 &       &  1 &165 &    &    &    & \\
 Ba\,II       &    3.78 & 0.47  &  3 &627 &2.06&    &  1 & 212\\
 La\,II       &    1.55 & 0.44  &  6 &116 &1.04&0.08&  4 & 20\\
 Ce\,II       &    1.53 & 0.16  &  5 & 83 &    &    &    &  \\
 Pr\,II       &    0.61 &       &  1 & 19 &    &    &    & \\
 Nd\,II       &    1.73 & 0.31  & 12 &102 &0.84&0.08&  4 & 8\\
 Eu\,II       &    0.01 & 0.04  &  2 & 20 &0.44&0.09&  3 & 20\\
\hline
\end{tabular}
\end{table*}

 A  lot  of  absorption lines of different elements (CNO-elements,
 light  metals,  iron  group  elements,  Ce,  Nd,  Eu)  have  been
 reliably   measured   in  the  spectrum  of  IRAS\,04296.  It  is
 important  that  we have not found any dependence of the abundances of
 these   species   on the equivalent  width  or on the excitation
 potential. Therefore  the  microturbulent velocity does not vary
 between different chemical elements.

 The  gf--values  for  most  of  the  spectral  lines used for the
 abundance  calculations  were  taken  from  the list used by Luck~(1991).  
The  S  and  CNO-abundances were determined by using the
 gf--data  from Waelkens et al.~(1991) and Giridhar et al.~(1994).
 The  list  of  lines  with  the  adopted  gf--values,  excitation
 potentials  of  the lower level and equivalent widths we measured
 for    the   object   IRAS\,04296   are   available   by   e-mail
 (valenta@alba.sao.ru).

 To  verify  the  method  of  analysis  we  observed with the same
 spectral  device  the  normal  supergiant  ${\alpha}$\,Per.
 The same procedures for processing and the same list of lines
 were used for analysis of the ${\alpha}$\,Per spectrum.
 This supergiant, whose parameters,  ${\rm  T_{eff}=6500\,K}$,
 ${\rm log\,g=1.2}$, ${\rm \xi_t=4.7\,km/sec}$  are very close to
 the object studied, is very convenient as a standard for the method
 testing because  of  its  membership  in  the young open cluster
 ${\rm   \alpha   Per}$   which  has  solar  chemical  composition
 (Klochkova \&  Panchuk~1985;  Boesgaard~1989).  Using its membership
 of  this cluster, we may  predict that
 ${\rm  \alpha\,Per}$  also  has normal solar chemical composition
 (aside  from  the expected nonsolar CNO triad abundances relative
 to iron). As it is shown in Table\,1 ${\rm \alpha Per}$ has indeed
 the abundances of chemical elements close to solar ones, except for
 CNO  and  several elements whose abundances are calculated with a
 large uncertainity due to a small number of spectral lines used.

\section{Spectral energy distribution}

 Details  of  the computer code used for solution of the radiative
 transfer   in  dusty  envelopes  can  be  found  in  Szczerba  et
 al.~(1997). In brief: the frequency--dependent radiative transfer
 equation   is   solved   for   a   dust   under   assumption   of
 spherically--symmetric  geometry for its distribution taking into
 account  particle  size  distribution and quantum heating effects
 for the very small dust particles.

 The  modelled  source is certainly C--rich (see Omont et al.~1995).
 Therefore, for modelling of its spectral
 energy  distribution  (SED)  we assumed that dust is composed of:
 policyclic aromatic hydrocarbons (PAH) for dust sizes $a$ between
 5   and   10\,\r{AA}   (see  Szczerba  et  al.~(1997)  for  details
 concerning  PAH  properties), amorphous carbon grains (of AC type
 from  Rouleau  \&  Martin~1991) for $a$\,$>$\,50\,\r{AA}, and dust
 with  an  opacity  obtained  from  averaging  of  the  absorption
 efficiences  for  PAH  and AC grains according to the formula: $$
 Q_{{\rm abs},\,\nu}\,=\,
    {\rm f}\,\cdot\,Q_{{\rm abs},\,\nu}^{{\rm PAH}}(a)\,
 +\,(1-{\rm  f})\,\cdot\,Q_{{\rm  abs},\,\nu}^{\rm  AC}(a),$$  for
 grain  sizes  between  10  and  50\,\r{AA}.  Here:  f\,=\,1 for $a$\,=\,
 10\,\r{AA}  and  f\,=\,0 for $a$\,=\,50\,\r{AA}. Dust with opacity values
 constructed  in  this way allow us to use a continous distribution
 of  dust  grain  sizes  and  fill  the  gap between properties of
 carbon--bearing molecules and small carbon grains.

 The  ${\rm  21\,\mu}$m  feature  was approximated by a gaussian with
 parameters   determined   from  modelling  of  IRAS\,07134$+$1005
 (centre  wavelength equal to ${\rm 20.6\,\mu}$m, and width of
 ${\rm  1.5\,\mu}$m)  which has the strongest
 feature  among the known ${\rm 21\,\mu}$m sources. In the case of
 ${\rm  30\,\mu}$m band we used the addition of two half--gausians
 with  the same strength and different width. Initial fit was done
 to  IRAS  22272$+$5435  and  its parameters were: width for short
 wavelength  side  $\sigma_{\rm  L}\,=\,{\rm  4\,\mu}$m, width for
 long  wavelength  side ${\sigma_{\rm R}\,=\,9\,\mu}$m and central
 wavelength  ${\rm  27.2\,\mu}$m  (see  Szczerba et al.~1997). For
 modelling  of  IRAS\,04296  we  have reduced the strength of this
 feature  by  50\%.  Superposition  of the 21 and ${\rm 30\,\mu}$m
 features  was  added to the absorption properties of amorphous carbon
 in order to construct an empirical opacity function (EOF).

\begin{figure}
\resizebox{\hsize}{!}{\includegraphics{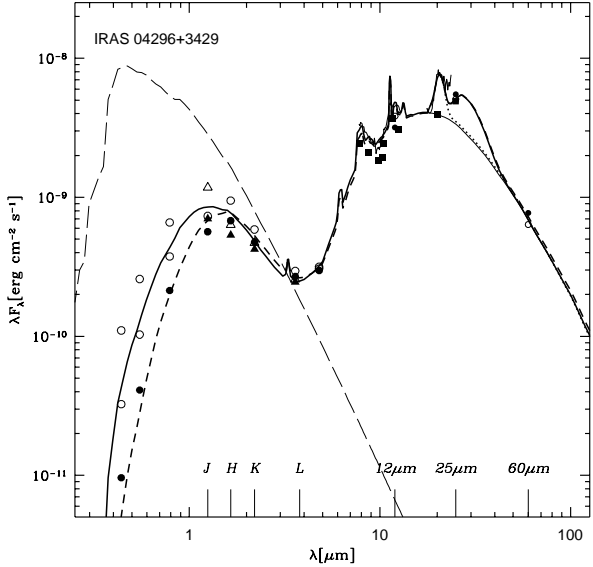}}
   \caption[]
 {Fit to the spectral energy distribution of IRAS\,04296\,+\,3429 obtained
with empirical opacity function, taking into account
quantum heating effects (heavy solid line) and assuming that
star radiates according to a model stellar atmosphere calculations for
log\,g\,=\,0.5 and $\rm T_{eff}$\,=\,6500 K (thin long--dashed line).
The heavy dashed line shows the fit obtained for star with
$\rm T_{eff}$\,=\,6000 K.  The thin solid line underlying 21
and ${\rm 30\,\mu}$m features represents the estimated
continuum level}
 \label{IRAS04296}
 \end{figure}

 In  Fig.\,\ref{IRAS04296} the best fit obtained from the solution
 of  the  radiative  transfer  problem  including  quantum heating
 effects  for  the PAH grains is shown together with observational
 data  which will be described in detail elsewhere. Note, however,
 that  we  present  also two sets of photometry (from B to M band)
 corrected for interstellar extinction (open symbols)
 according   to   the   average  extinction  law  of  Cardelli  et
 al.~(1989),  assuming  that  total  extinction at V is 1.0 or 2.0
 magnitudes  and  plotting  only the smallest and largest value of
 corrected fluxes at given band. This  estimate of the total extinction 
range can be inferred from
 the analysis of data presented by Burstein \& Heiles~(1982).
%

\begin{table}
\caption[]{Model parameters for IRAS\,04296+3429.
   Precise meaning of the symbols used can be found in Szczerba et
al.~(1997)}
\begin{tabular}{ l l }
\hline
\noalign{\smallskip}
      parameter  & value \\
\noalign{\smallskip}
\hline
\noalign{\smallskip}
   $\rm T_{eff}$                     &  6500\,\,K \\

   log (L$_{\rm star}$\,[L$_{\odot}$])       &  3.92    \\

   $d$                                &  5.4\,\,kpc \\

                                      &            \\

   $R_{\rm out}$                      &  0.5\,\,pc  \\

   $V_{\rm exp}$                    &  12\,km\,s$^{-1}$ \\

                                    &       \\

   $R_{\rm in}$(hot dust shell)      &  6.4\,10$^{-4}$\,\,pc \\

   $\overline{T}_{\rm d}$[R$_{\rm in}$(hot dust shell)] &  870\,\,K \\

   $\rho_{\rm gas}$ (hot dust shell)  &  $\sim$\,r$^{-2.0}$ \\

   $\dot{M}_{\rm post-AGB}$      &  4.0\,10$^{-7}$\,M$_{\odot}\,{\rm yr}^{-1}$\\

                                  &            \\

   $R_{\rm in}$(main shell)           &  7.06\,10$^{-3}$\,\,pc \\

   $\overline{T}_{\rm d}$[R$_{\rm in}$(main shell)]    &  270\,\,K  \\

   $\rho_{\rm gas}$ (main shell)      &  $\sim$\,r$^{-2.6}$ \\

   $\dot{M}_{\rm AGB}^{\rm min}$  &  1.70\,10$^{-5}$\,M$_{\odot}\,{\rm yr}^{-1}$ \\

   $\dot{M}_{\rm AGB}^{\rm max}$        &  2.19\,10$^{-4}$\,M$_{\odot}\,{\rm yr}^{-1}$ \\

                                  &            \\

   $a_{-}$                           &  5\r{AA} \\

   $a_{+}$                           &  0.25\,$\mu$m \\

   $p$                               &  3.5      \\

                                    &       \\

  $t_{\rm dyn}$                       &  575\,\,yr \\

   $M_{\rm dust}$                       & 0.0071\,M$_{\odot}$ \\
\noalign{\smallskip}
\hline
\end{tabular}
\end{table}

 The  best  fit to the spectral energy distribution of IRAS\,04296
 is   shown   by  heavy  solid  line  (see  Table\,2  for  details
 concerning  parameters of the model). Our modelling procedure was
 such  that  we tried to get fits to the SED which fall in between the
 extinction  corrected  fluxes.  In  this  way, we have taken into
 account  not  only the effect of the circumstellar extinction but
 also  of  interstellar  extinction.  The  thin  long--dashed line
 represents  the input energy distribution of the central star for
 log\,g\,=\,0.5   and   $\rm T_{eff}$\,=\,6500\,K   according   to   model
 atmosphere  calculations  of  Kurucz (private communication). The
 heavy  short--dashed  line  shows the fit which was obtained with
 the  same  assumptions  but changing the effective temperature of
 the  star  to 6000\,K. As one can immediately see in the IR range
 of  the  spectrum  the  quality  of  the  fits  are very similar.
 However,  in  the  optical  and  ultraviolet  (UV)  part  of  the
 spectrum  the  fit  assuming  $\rm T_{eff}$\,=\,6000\,K  is  not  able to
 explain  extinction  corrected data. In consequence, we are quite
 convinced  that  our  estimation of $\rm T_{eff}$ for IRAS\,04296 close
 to  6500\,K  is  reasonable  and,  what  is  even more important,
 agrees  pretty  well with the spectroscopic estimation (6300\,K).
 Note  that  spectral  type  of  this source was found to be G0~Ia
 from  the low resolution spectrum (Hrivnak~1995) which implies an
 effective  temperature of around 5500\,K for the star if we asume
 that  the  same relationship applies for post--AGB supergiants as
 for  ``normal''  ones  (see  Schmidt--Kaler~1982). For such a low
 temperature  we  were  not able to fit even the reddenned data in
 the UV.

 The  thin  solid  line  in  the wavelength range from about 18 to
 ${\rm 48\,\mu}$m represents the model continuum level found after
 solution  of  radiative  transfer equation for dust without EOF using
 the parameters  as  in  Tab.\,2  while  keeping  the  dust
 temperature  (or  probability  distribution  of dust temperature)
 the  same  as  for the case of dust with EOF. Taking into account
 the  estimated continuum level and assuming that 21\,$\mu$m feature
 extends  from  18 to 22\,$\mu$m we estimate the energy emitted in 21\,$\mu$m
  band  as  about 5.7 \% of the total IR flux (251\,L$_{\odot}$ for
 $\lambda$'s  from 5 to 300\,$\mu$m assuming a distance to the source
 of  1  kpc).  With the dotted line for wavelengths longer than 18\,$\mu$m
  we  present  the  fit  which  was obtained using an opacity
 function  with  the  EOF  for only 21\,$\mu$m component. It is clear
 that  such  fit is not able to explain IRAS photometry at 25\,$\mu$m.
 Our  recent  {\it ISO} observations show that this source is also a
 30\,$\mu$m  emitter. In the forthcoming paper (Szczerba et al.~1999)
we will discuss this finding in detail.

\section{Discussion}

 As it is  shown  in Table\,1, the metallicity for IRAS\,04296
 is significantly decreased relative to the solar  value: the average
 abundance for the elements of the iron-group   with  respect  to  the  Sun
 is  ${\rm [(Ti,V,Cr,Fe)/H]_{\odot}\,=\,-0.9}$ with the standard
 deviation ${\rm \sigma = 0.2}$.

 Recently  Decin  et al.~(1998), using high resolution spectra and
 model  atmospheres  method,  calculated abundances of 14 chemical
 elements  in  the  IRAS\,04296  atmosphere.  Their results are in
 qualitative  agreement  with these ones presented here, but there
 are  some significant differences. Decin et al.~(1998) calculated
 chemical  composition  of this object assuming $\rm T_{eff}$\,=\,7000\,K,
 log\,g\,=\,1.0,  $\xi_t$\,=\,4\,km/s,  rather different from
 the  model  atmospheres  parameters found in this work. It should
 be  noted  that  we  estimated  the  effective temperature by two
 independent  methods,  and  it  is  worth  stressing that we have
 obtained   consistent   values   of  the  effective  temperature:
 $\rm T_{eff}$\,=\,6300\,K  from numerous Fe\,I, Fe\,II spectral lines and
 $\rm T_{eff}$\,=\,6500\,K   from   modelling   of   the  spectral  energy
 distribution   of   this  source.  The  difference  in  effective
 temperature  between  Decin  et  al.~(1998)  and our estimation
 (${\Delta}$\,$\rm T_{eff}$\,=\,700\,K)   is  able  to  explain  different
 metallicities  estimated  by Decin et al.~(1998) and by us (${\rm
 \Delta\,log\,\epsilon(Fe)\approx\,0.2}$).  The  same  is true for
 the   case   of   the   rare-earth   element   abundances:  large
 differences,  about  1\,dex,  in the values could be explained by
 differences in model atmosphere parameters.

 Let  us  consider  now  in  more  detail the peculiarities in the
 chemical   composition  of  the  object.  For  this  purpose,  in
 Table\,3 we present the logarithmic differences
 $${{\rm [X/Fe]_{\odot}\,=\,[log\,\epsilon(X) - log\,\epsilon (Fe)]_{\star}
- [log\,\epsilon(X) - log \,\epsilon  (Fe)]_{\odot}}}$$
 between  chemical  compositions  of different objects and the Sun
 (solar  abundaces  from  Grevesse  et  al.~(1996)): in the second
 column  for IRAS\,04296, in the third column for IRAS\,07134+1005
 (hereafter  IRAS\,07134  -  asscociated with the peculiar F--type
 supergiant  HD\,56126)  and  in  the  fourth  one  for  the  star
 ROA\,24.  The  objects  are  similar  from point of view of their
 atmospheric  parameters  ($\rm T_{eff}$,  log\,g)  and relative chemical
 composition.   It  should  be  noted  that  the  metal  deficient
 supergiant  ROA\,24  (Fehrenbach's  star) belongs to the globular
 cluster  ${\rm \omega\,Cen}$ and could be considered as a typical
 {\it halo} object in the post-AGB evolution stage.

\begin{table}
\caption{Relative abundances of chemical
         elements for IRAS\,04296+3429 in comparison to
         related PPNe. The data by Grevesse \& Noels~(1996) are adopted
         for solar abundances.}
\begin{tabular}{lcccc}
&&&& \\
\hline\LARGE
 & IRAS\,04296+3429 & IRAS\,07134+1005$^a$ & ROA\,24$^b$  \\
  &${\rm [Fe/H]_{\sun}\,=\,-0.84}$  &-1.00 &-1.77  \\[5pt]
\hline
Element &\multicolumn{3}{c}{${\rm [X/Fe]_{\odot}}$}\\[5pt]
\hline
 Li\,I  & ${\rm \ge +0.23 }$ &      &      \\
 C\,I   &  +0.84 &+1.08 &+0.67 \\
 N\,I   &  +0.83 &+1.03 &+1.02 \\
 O\,I   &  +0.19 &+0.63 &+1.01 \\
 Na\,I  &  +0.42 &+0.54 &+0.71 \\
 Mg\,I  &        &+0.97 &+0.31 \\
 Mg\,II &  +1.34 &      &+0.09 \\
 Al\,I  &  +1.03 &+1.48 &      \\
 Si\,I  &  +0.58 &+0.95 &+0.80 \\
 Si\,II &  +0.26 &      &+1.03 \\
 S\,I   &  +0.43 &+0.63 &      \\
 Ca\,I  &  +0.19 &+0.45 &+0.60 \\
 Sc\,II &  +0.18 &-0.07 &-0.13 \\
 Ti\,II &  -0.27 &      &+0.33 \\
 V\,II  &  +0.10 &-0.03 &+0.15 \\
 Cr\,II &  +0.11 &      &+0.65 \\
 Cu\,I  &  +0.24 &+1.03 &-0.01 \\
 Zn\,I  &  +0.08 &      &      \\
 Y\,II  &  +1.20 &+1.70 &+0.37 \\
 Zr\,II &  +0.62 &      &      \\
 Ba\,II &  +2.49 &+0.99 &+0.96 \\
 La\,II &  +1.17 &+1.59 &+0.54 \\
 Ce\,II &  +0.82 &      &+1.60  \\
 Pr\,II &  +0.74 &      &      \\
 Nd\,II &  +1.07 &+1.30 &+0.67 \\
 Eu\,II &  +0.34 &+1.06 &+0.25 \\
\hline
\multicolumn{4}{l}{a -- Klochkova~(1995),} \\
\multicolumn{4}{l}{b -- Gonzalez \& Wallerstein~(1992).}\\
\end{tabular}
\end{table}

The carbon overabundance ${{\rm   [C/Fe]_\odot\,=\,+0.8}}$
  (revealed  from  intensities  of  21  absorption  lines
  with the standard deviation ${\rm \sigma = 0.46}$) and  the
  enhancement  of  nitrogen ${{\rm [N/Fe]_\odot\,=\,+0.8}}$ (from 4
  lines, ${\rm \sigma = 0.10}$) suggest that IRAS\,04296  underwent the
  third dredge--up episode.

 The  oxygen  content  based  on intensity of 3 weak lines
 near  ${\rm \lambda \approx }$ 6155\,\r{AA} is determined with a small
 internal error.

 From  the  Fe--deficiency  and CNO abundances (${\rm C/O\,>\,2}$)
 we  can  conclude  that  IRAS\,04296  is a low mass object in
 advanced   stage of evolution.  For an  {\it  unevolved}
 metal--deficient object (with  ${\rm  [Fe/H]_{\odot} \approx -0.9}$)
 the  average  value of ${\rm  [C/Fe]_{\odot}}$ is only about
 -0.2 (Tomkin  et al.~1995), the average value of ${\rm [N/Fe]_{\odot}}$
 is ${\rm \approx  0}$ (Wheeler  et al.~1989, Timmes et al.~1995)
 and the average value of ${\rm [O/Fe]_{\odot}}$ is ${\approx +0.5}$
 (Wheeler  et  al.~1989, Timmes  et al.~1995, Klochkova \& Panchuk~1996b).
 The atmospheres of the post-AGB stars IRAS\,07134 and ROA\,24 are also
 overabundant in both carbon and nitrogen. Note however, that for
 most  of  the  PPNe  candidates  studied, strong relative changes
 between   elements  of  the  CNO--group  are  observed  (Luck  et
 al.~1983;  Lambert  et  al.~1988;  Klochkova~1995;  Za\v{c}s  et
 al.~1995,  1996;  van  Winckel  et  al.~1996a, 1996b; van Winckel~1997).

 The abundances of some light metals (Na, Al, Mg, Si, Ca) are enhanced
 for all three stars. The average value for these elements is
 ${\rm  [X/Fe]_{\odot}\,=\,+0.6}$ for IRAS\,04296; +0.9 for IRAS\,07134 and
 +0.6 for ROA\,24, with the standard deviations: ${\rm \sigma\,=\,0.4}$,
 0.4 and 0.36, respectively.

 We  did  not  still  include the K\,I abundance into our results,
 since  we  suspect  that  the  equivalent  width of its line near
 ${\rm  \lambda}$  7699\,\r{AA}  could be significantly distorted due
 to circumstellar and interstellar components.

 The  iron--group element zinc is the most important for determination
 of real (initial) value of the metallicity of a star since, firstly,
 its abundance follows that of iron in  a wide [Fe/H]$_{\odot}$ interval
 (Sneden \& Crocker~1988; Wheeler et al.~1989, Sneden et al.~1991) and,
 secondly, zinc having a low condensation temperature is not depleted by
 selective  separation processes onto dust grains (Bond~1992).
 A close to solar abundance of Zn relative to iron
 (${\rm  [Zn/Fe]_{\odot}\,=\,+0.1}$) permits us to conclude about the
 inefficiency of the selective separation processes in the IRAS\,04296
 envelope. This conclusion is based also on an absence of overdeficiency
 of light depleted elements (Ca, Sc).
 Besides, the relative  abundance (${\rm  [S/Fe]_{\odot}\,=\,+0.4}$
 with the standard deviation  ${\rm \sigma = 0.21}$) of S, a chemical
 element which is not depleted by dust--gas separation, for IRAS\,04296
 is close to the value for unevolved metal-deficient  dwarfs
 (Fran\c{c}ois~1987, Timmes et al.~1995). This futher confirms
 the lack of selective separation in the envelope of the object studied.

 Individual  abundances of the heavy s-process metals Y and Zr are
 determined  with  a  relatively  large error because of the small
 number  of  lines  measured.  However,  the  average  value ${\rm
 [X/Fe]_{\odot}\,=\,+0.9}$   for   Y   and   Zr  is  sufficiently
 reliable.  In  addition, the abundance of heavy s-process element
 Ba    (${\rm   [Ba/H]_{\odot}\,=\,+2.5}$)   derived   from   the
 equivalent   width   of  strong  lines  could  be  altered  by  a
 systematic  error  due  to the complexity of the outer regions of
 the  stellar  atmopshere  as  discussed  above.  Nevertheless, we
 conclude that there is a Ba excess.

 The  abundance of lanthanides (La, Ce, Pr, Nd)  are strongly enhanced relative
 to iron for the objects from Table\,3. For these heavy metals the
 average  value is ${{\rm [X/Fe]_{\odot}\,=\,+1.0,\,+1.4,\,+0.9}}$
 for IRAS\,04296, IRAS\,07134 and ROA\,24, respectively, with the standard
 deviations 0.2 and 0.6 for  IRAS\,04296 and ROA\,24. Moreover,
 for  all  these  objects  we see the overabundance of Eu which is
 predominantly produced by the r--process.

 Excess  of  s-process  elements has been reliably found up to now
 in   three   objects   investigated   at   the   6\,m  telescope:
 IRAS\,04296+3429,    IRAS\,07134+1005    and    IRAS\,22272+5435.
 Besides,  similar  conclusions have appeared for another four PPN
 candidates  (and  for  one  object  in  common):  HD\,158616 (van
 Winckel   et   al.~1995);   IRAS\,19500-1709\,=\,HD\,187885  (van
 Winckel~1997);    IRAS\,05341+0852    (Reddy   et   al.~1997);
 IRAS\,22223+4327  and  IRAS\,04296+3429  (Decin  et al.~1998). In
 atmospheres  of  most PPN candidates overdeficiency (with respect
 to  their  metallicity)  of  heavy  nuclei  is generally observed
 (Klochkova~1995;  van Winckel et al.~1996a, 1996b; Klochkova \&
 Panchuk~1996a;  van  Winckel~1997),  whose  existence  in  the
 atmospheres  of post--AGB low--mass supergiants has not yet found
 a clear explanation.

 In consequence, we  can  state that chemical abundances pattern
 for  the  source  IRAS\,04296  is  related  to  its old galactic
 population  membership  and  dredge--up of matter enriched by the
 nucleosynthesis products. It may be part of the old disk population.

 As  has  been  concluded  already  by Decin et al.~(1998) all the
 post--AGB  candidates  mentioned  above  (only  these, up to now,
 show  an  s--process  element  enhancement!)  belong to the small
 group  of PPNe (Kwok et al.~1989; Kwok et al.~1995) which have in
 their  IR  spectrum  an unidentified emission band at about ${\rm
 21\,\mu}$m.  This  feature  is  neither found   in the
 spectra  of  their  predecessors, AGB stars, nor in the spectra of
 PNe.  Note,  once  more,  that the search by means of the {\it ISO} for the
 new  21\,$\mu$m  emitters  among  candidates selected by Henning et
 al.~(1996) failed (Henning,  private  communication).  
 As  has  been  stated in the
 papers  by  Kwok  et  al.~(1989, 1995), the objects whose spectra
 contain  the  ${\rm  21\,\mu}$m  band  are carbon-rich stars. Our
 investigations  based on the spectra from the 6\,m telescope, for
 IRAS\,07134   (Klochkova~1995),  IRAS\,22272+5435  (Za\v{c}s  et
 al.~1995)  and  IRAS\,04296  (Klochkova  et al.~1997b), confirmed
 that  ${\rm  C/O>1}$  for  all  of  them.  In  this  context, the
 conclusion  that  the carrier of the 21\,$\mu$m band is related to
 C  is natural. For example, Buss et al.~(1990) have supposed that
 this  feature  may  be  caused  by polycyclic
 aromatic  hydrocarbons.  On  the  other  hand,  Goebel~(1993) has
 identified  the  ${\rm  21\,\mu}$m band with the vibrational band
 of  the  SiS$_2$  molecule,  the  presence of which is consistent
 with the temperature in the envelope.

 Taking   into   account   the   available   results  on  chemical
 composition  for  subclass  of  PPNe with the 21\,$\mu$m feature:
 IRAS\,07134+1005  (Parthasarathy  et  al.~1992,  Klochkova~1995),
 IRAS\,22272+5435  (Za\v{c}s  et  al.~1995),  IRAS\,19500-1709 (van
 Winckel  et  al.~1996a),  IRAS\,05341+0852  (Reddy  et al.~1997),
 IRAS\,22223+4327  (Decin  et al.~1998), and IRAS\,04296 (Decin et
 al.~1998;  Klochkova  et  al.~1997b;  this paper) we see that the
 carbon-rich  atmospheres  of  these  objects are also enriched by
 s--process  elements.  It  is  evident  that  there  is  a strong
 correlation  between  presence  of  the 21\,$\mu$m feature, ${\rm
 C_2}$,  ${\rm C_3}$ molecular bands, and excess of the s--process
 elements.  Decin  et  al.~(1998)  were  the first who pointed out
 this  relationship.  What  is  even more important, an excess of
 s--process  elements  was  not  found for a number of IRAS sources
 with altered CNO-content but without the 21\,$\mu$m feature (some
 of  which  are  oxygen-rich stars rather than carbon-rich stars):
 IRAS\,06338$+$5333  (Luck  \&  Bond~1984;  Bond  \&  Luck~1987),
 IRAS\,07331$+$0021  (Luck  \&  Bond~1989;  Klochkova  \& Panchuk~1996a),
   IRAS\,09276$+$4454   (Klochkova   \&  Mishenina~1998),
 IRAS\,12175$-$5338  (van  Winckel~1997), IRAS 12538$-$2611 (Luck
 et  al.~1983;  Klochkova \& Panchuk~1988b; Giridhar et al.~1997),
 IRAS\,15039$-$4806      (van      Winckel      et     al.~1996b),
 IRAS\,17436$+$5003   (Klochkova   \&   Panchuk~1988a;  Luck  et
 al.~1990;  Klochkova~1998), IRAS\,18095$+$2704 (Klochkova~1995),
 and  IRAS\,19114$+$0002  (Za\v{c}s  et al.~1996, Klochkova~1998).
 Therefore,  it  seems  that  carrier  of  21\,$\mu$m feature is {\it
 strongly}  related  to  the  whole  chemical  composition pattern
 typical  for the third dredge--up (excess of s-process elements),
 and  not  only  to  the  C-richness of the photosphere.

 That  21\,$\mu$m feature is not observed around AGB--stars showing
 s--process   elements   could   be   explained  by  the  physical
 conditions  which  are  inappropriate  for the excitation of this
 band,  while  its  non--presence  in  planetary  nebulae may be a
 result of carrier destruction by the highly energetic photons.

\section{Conclusions}

 We  conlude  that  IRAS\,04296$+$3429  is  a PPN candidate with a
 chemical    composition    which   coincides   with   theoretical
 predictions  for  the  post--AGB  objects:  very  large excess of
 carbon and nitrogen are  revealed.  Moreover,  the  real
 excess  (relative to iron) of heavy metals Y, Zr, Ba, La, Ce, Pr,
 Nd   synthesized  by  the  neutronization  process  indicates  an
 effective  third  dredge--up  and further confirms IRAS\,04296 to
 be in the advanced post-AGB evolution stage.

 The  emission  of  ${\rm  C_2}$ molecular lines discovered in the
 spectra  of IRAS\,04296 and its similarity to the emission of the
 Hale--Bopp  comet allow us to suggest that in both cases the same
 mechanism  (the  resonance  fluorescence)  is responsible for the
 observed features.

 Several  strong  absorptional  features  whose positions coincide
 with  known diffuse interstellar bands (DIBs) are found in the
 spectrum of IRAS\,04296.

 In  addition,  from  the  SED  modelling  of  the spectral energy
 distribution  we showed that 25\,$\mu$m flux cannont be explained
 without  assumption  that  the  30\,$\mu$m  emission  feature  is
 present  in  this source. Our recent {\it ISO} observation has detected
 the 30\,$\mu$m band in this source.

\begin{acknowledgements}
 We are much indebted to the
 referee Hans van Winckel for critical reading the manuscript and
 valuable advice.

 This  work  has  been supported by project 1.4.1.1 of the Russian
 Federal   Program   ``Astronomy''  and  grant 
 2.P03D.002.13  of  the  Polish  State  Committee  for  Scientific
 Research.  One  of us (R.Sz.) gratefully acknowledges the support
 from the Canadian Institute of Theoretical Astrophysics.
\end{acknowledgements}

\end{document}